\documentclass{pasj00}

\begin{document}
\SetRunningHead{Hansteen et al.}{Connecting SOT and EIS}
\Received{2000/12/31}
\Accepted{2001/01/01}

\title{On connecting the dynamics of the chromosphere and transition region with Hinode SOT and EIS.Ê}

\author{Viggo H.\textsc{Hansteen}\altaffilmark{1,2}
        Bart \textsc{De Pontieu}\altaffilmark{2}
        Mats {\sc Carlsson}\altaffilmark{1}
        Scott {\sc McIntosh}\altaffilmark{3,4}}
\email{Viggo.Hansteen@astro.uio.no}
\author{Tetsuya {\sc Watanabe}\altaffilmark{5}
        Harry {\sc Warren}\altaffilmark{6}
        Louise {\sc Harra}\altaffilmark{7}
        Hirohisa {\sc Hara}\altaffilmark{5}}
\author{Theodore D. {\sc Tarbell}\altaffilmark{2}
        Dick {\sc Shine}\altaffilmark{2}
        Alan {\sc Title}\altaffilmark{2}
        Carolus J. {\sc Schrijver}\altaffilmark{2}}
\author{Saku {\sc Tsuneta}\altaffilmark{5}
        Yukio {\sc Katsukawa}\altaffilmark{5}
        Kiyoshi {\sc Ichimoto}\altaffilmark{5}
        Yoshinori {\sc Suematsu}\altaffilmark{5}
        Toshifumi {\sc Shimizu}\altaffilmark{8} }
\altaffiltext{1}{Institute of Theoretical Astrophysics, University of Oslo, PB 1029 Blindern, 0315 Oslo Norway}
\altaffiltext{2}{Lockheed Martin Solar and Astrophysics Laboratory, Palo Alto, CA 94304, USA}
\altaffiltext{3}{Department of Space Studies, Southwest Research Insititute, 1050 Walnut St, Suite 400, Boulder, CO 80302, USA}
\altaffiltext{4}{High Altitude Observatory, National Center for Atmospheric Research, PO Box 3000, Boulder, CO 80307, USA}
\altaffiltext{5}{National Astronomical Observatory of Japan, Mitaka, Tokyo, 181-8588, Japan}
\altaffiltext{6}{Space Science Division, Naval Research Laboratory, Washington DC, USA}
\altaffiltext{7}{Mullard Space Science Laboratory, University College London, UK}
\altaffiltext{8}{ISAS/JAXA, Sagamihara, Kanagawa, 229-8510, Japan}

%

\KeyWords{Sun: chromosphere -- Sun:transition region -- Sun:corona -- Sun: UV radiation
          -- Sun: flux emergence} 

\maketitle

\begin{abstract}
We use coordinated Hinode SOT/EIS observations that include
high-resolution magnetograms, chromospheric and TR imaging and
TR/coronal spectra in a first test to study how the dynamics of the TR are driven by the
highly dynamic photospheric magnetic fields and the ubiquitous
chromospheric waves. Initial analysis shows that these connections are
quite subtle and require a combination of techniques including  magnetic 
field extrapolations, frequency-filtered time-series and comparisons with 
synthetic chromospheric and TR images from advanced 3D numerical simulations. As a first result, 
we find signatures of magnetic flux emergence as well as 3 and 5~mHz wave power
above regions of enhanced photospheric magnetic field in both chromospheric, transition 
region and coronal emission.
\end{abstract}

\section{Introduction}

The Hinode spacecraft, launched in late September 2006, is comprised of 
three scientific instruments \citep{Hinode-overview}: the Solar Optical Telescope (SOT), the 
Extreme ultraviolet Imaging Spectrograph (EIS), and the X-Ray Telescope
(XRT). The SOT is designed to produce high quality images and measurements 
of the magnetic field in various photospheric and chromospheric lines and continua
\citep{SOT-instrument}. 
The ultraviolet and X-ray instruments are constructed to extract information
from the outer solar atmosphere. In particular, EIS \citep{EIS-instrument} 
observes in two bands ($180 - 205$~{\AA} and $250 - 290$~{\AA}) that feature a 
number of coronal and some transition region emission lines. 

An explicitly stated primary scientific goal of the Hinode
mission is to map and understand the transport of mechanical energy 
flux between the lower lying layers of the Sun's atmosphere and the corona, as well 
as the relation between the structure of the photospheric magnetic field and coronal 
heating. 
Generally speaking there are two ways of transporting energy
from the convection zone and into the layers above: either by utilizing
waves, such as the 5-minute p-modes or chromospheric 3-minute
oscillations, or by converting the energy contained in the magnetic
field, stressed by photospheric flows and granular evolution, 
in some episodic fashion into heat in the chromosphere and/or
corona. Related questions are what role magnetic flux emergence plays in 
injecting energy into the chromosphere/corona and replenishing the previously 
existing field; and with what efficiency acoustic
and Alfv{\'e}n waves are generated and propagated through the chromosphere and
transition region. In this paper we report on preliminary 
simultaneous observations made with the SOT and EIS instruments to shed light
on these issues.

\section{Instrumental setup and Observations}

During a two-week period in February 2007 we observed several types of 
solar region including coronal hole, quiet Sun, network, and plage/small 
active regions both on the disk and towards the solar limb. A substantial
amount of time was spent following the progress of the small active region
NOAA 10942 as it transverses the disk from a solar position of roughly 
$-400$~arcsec to the west of disk center on February 20 to a position of
$900$~arcsec east, near the eastern limb on February 27. 

With SOT we obtained magnetograms measured in the Fe {\sc i} $630.2$~nm line
and Ca~{\sc ii} $396.8$~nm H-line images at high ($4-11$~s cadence) for the observations
described here. The Ca~{\sc ii} H-line filter used with SOT is fairly wide (0.22~nm) and
thus includes a significant fraction of line wing in addition to line core.
We have collected similar image series in the G-band and in the blue continuum channel of 
SOT's broad band filtergraph but those observations are not described here. 

The EIS instrument also supports a number of observing modes. We report on 
observations made with the 40~arcsec wide slot
that forms a 40~arcsec wide image of the Sun in particular (strong) emission 
lines on the detector. This observing mode has the advantage of allowing 
images to be recorded at relatively high cadence (30~s or so for the strongest 
lines), but incurs the cost of removing any easily derived velocity or 
line-width signal as well as a slightly worse spatial resolution that that achieved with
the 1~arcsec slit. We also report on raster observations made by stepping the
1~arcsec slit in 1~arcsec increments across the region of interest. With exposure times of 
30-60~s a typical raster takes 
of order an hour to complete. The spectral resolution of EIS is $0.00223$~nm;
roughly 27~km/s at 25.0~nm or 35~km/s at 19.0~nm. We have also 
collected several sit-and-stare time series with an immobile 1~arcsec slit, but will 
present results from these at a later date. 

In normal operation the Hinode spacecraft tracks a given solar feature, correcting 
for solar rotation. In principle, co-alignment between data sets obtained by the
various instruments becomes a question of finding the offsets between the instruments
and applying these to the collected data.
After the EIS coarse mirror move of January 24 2007, the EIS offset is some 3~arcsec
east and 50~arcsec south of the center of the SOT field of view. In actual practice finding
the correct offsets is not always straightforward. As will be emphasized in this paper
it is sometimes far from simple to identify corresponding solar features at photospheric/lower
chromospheric heights with those found in the transition region and corona, especially 
considering the mismatch in spatial resolution between the SOT and EIS instruments. 
When in serious doubt about alignment we found it helpful to co-align via TRACE observations
by identifying features found in TRACE Fe~{\sc ix/x} 17.1~nm and EIS Fe~{\sc xii}, 
as well as the TRACE~160.0~nm channel and the SOT Ca H-line.

It must also
be borne in mind that SOT has a correlation tracker that removes most of the
orbital variations on SOT pointing while EIS does not have such a device: Flexing 
due temperature variations of the EIS instrument  during an orbit can be as large as 
5~arcsec in the north/south direction and $2-3$~arcsec in the east/west direction. 
We have removed some of this 
variation by internally co-aligning the images in EIS time-series. 
We believe we have achieved a co-alignment that is generally better than
2~arcsec.

\section{Results}
\subsection{Quiet Sun with weak plage}

\begin{figure}
  \begin{center}
    \FigureFile(80mm,70mm){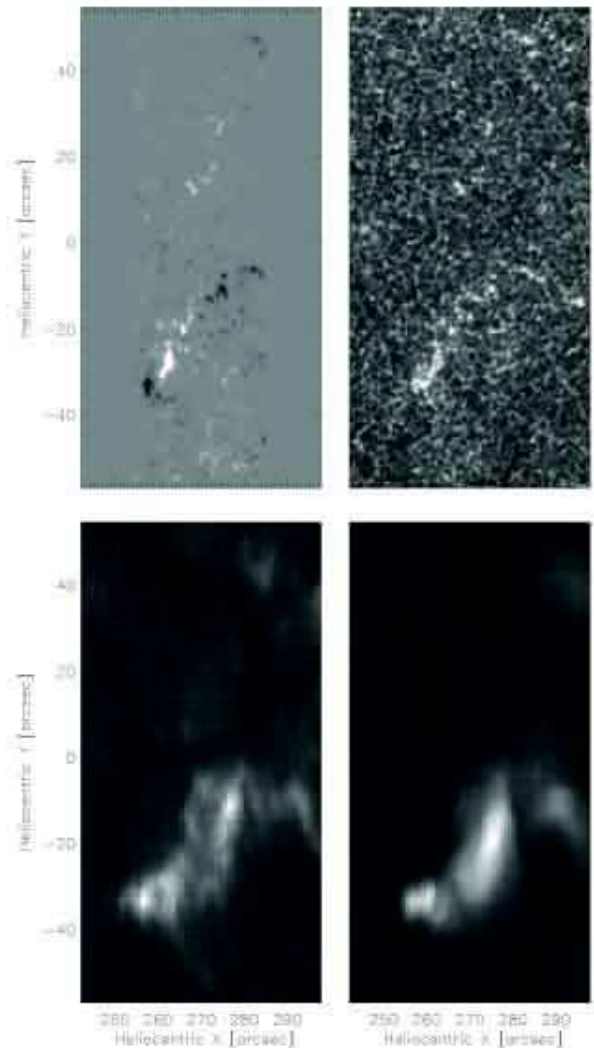}
  \end{center}
  \caption{Co-pointed SOT and EIS images of a quiet Sun region centered on heliocentric 
  co-ordinates (280,0)~arcsec obtained February 19, 2007. In the upper left panel we show 
  a Fe~{\sc i} 630.2~nm 
  magnetogram and in the upper right panel the Ca~{\sc ii} 396.8~nm H-line. In the lower panels
  the corresponding EIS raster images for the transition region He~{\sc ii} 25.6~nm (left) and the
  coronal Fe~{\sc xii} 19.5~nm (right) lines are shown.}\label{fig:intensity_19Feb2007}
\end{figure}

On February 19 2007, from roughly 11:30 to 17:30 we observed a quiet Sun
region that also contained weak plage of both polarities. In 
Figure~\ref{fig:intensity_19Feb2007} we show co-pointed individual frames from 
the SOT and EIS rasters that zoom in on the region observed with the SOT.

\begin{figure*}
  \begin{center}
    \FigureFile(150mm,150mm){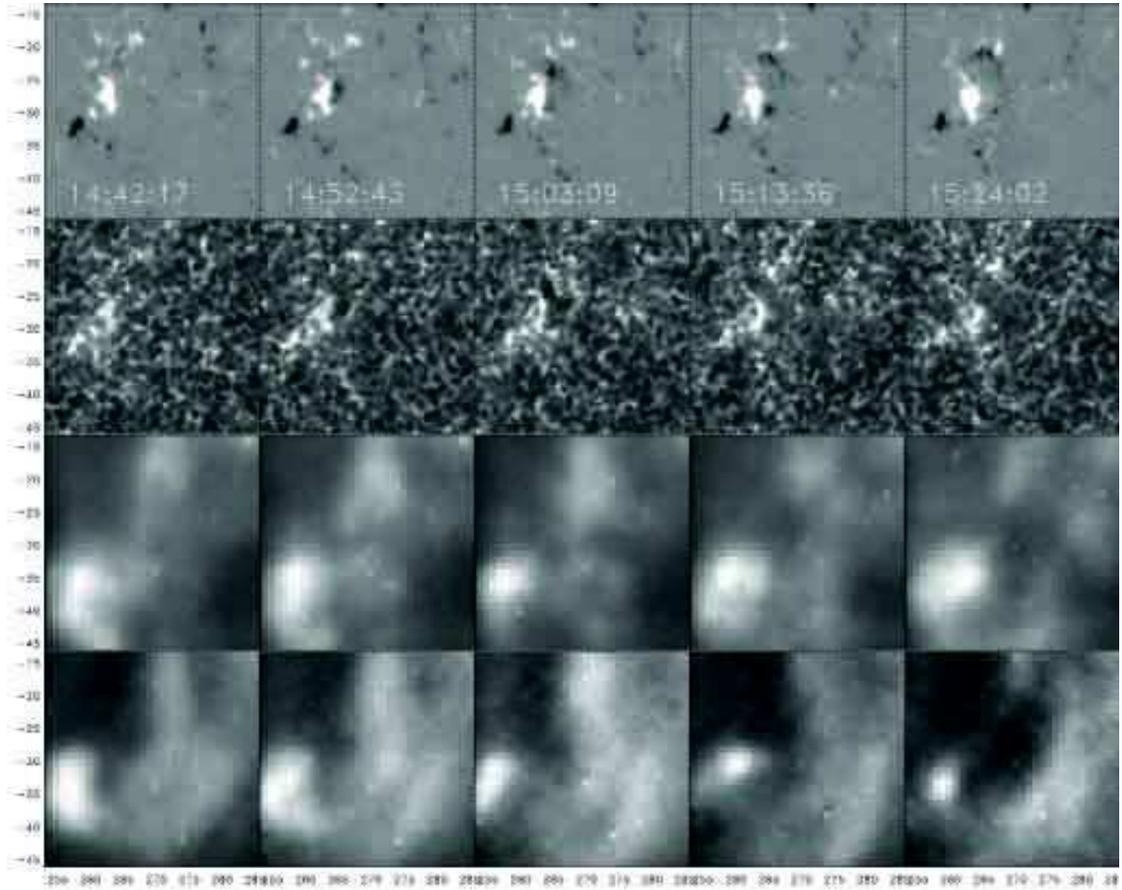}
  \end{center}
  \caption{Co-pointed SOT and EIS images of a quiet Sun region centered on heliocentric 
  co-ordinates (280,0)~arcsec obtained February 19, 2007. This set of images shows
  the time evolution of a cutout of the region shown in Figure~\ref{fig:intensity_19Feb2007}
  that illustrates the flux emergence that occurs from roughly 14:45~UT to 15:30~UT (at which
  time the EIS time series completed). The top panels show Fe~{\sc i} 630.2~nm magnetograms 
  separated by some 10~minute intervals. The three following rows of panels show the 
  chromospheric Ca~{\sc ii} 396.8~nm H-line, the transition region He~{\sc ii} 25.6~nm line and
  the coronal Fe~{\sc xii} 19.5~nm line at the same instants in time.}
  \label{fig:emergence_19Feb2007}
\end{figure*}

In the upper left panel we show an image from the SOT magnetogram movie. 
The field consists of several small magnetic elements of both polarities
spread across the image. In addition, we find a stronger plage region 
stretching from (255,-35) to (280,-5) in heliocentric coordinates. 
There is also some weaker plage centered at (270,15) and 
(283,45). In the movie the magnetic field, churned by the motions 
of the photosphere, shows motions of the individual flux elements that 
presumably correspond to G-point bright points as well as the bright 
points seen in the Ca~{\sc ii} H-line images. Flux elements of opposite 
polarity sometimes meet and merge, disappearing from view, at other times 
flux elements appear and move apart. In the weak plage regions of greater 
magnetic field concentration the field seems less responsive to photospheric 
motions, or alternately photospheric motions are suppressed by the presence 
of the field. The general topology of the plage is unchanged during the six hours
this region of the Sun was observed. We did however, observe an incidence
of flux emergence in the central portion of the plage region; negative
(black) field appears at (265,-25) just to the right of the large collection of
positive (white) field present in figure~\ref{fig:intensity_19Feb2007}, 
rapidly splitting in two with one collection of field moving northeast and 
the other more or less staying in place. This new flux is largely dissipated at
the end of the time-series roughly one hour after it first appears. Detailed 
images of the time evolution of the flux emergence are shown in 
Figure~\ref{fig:emergence_19Feb2007}.

The magnetic topology outlined by the small plage region is evident in
all three of the wavelength bands shown in Figure~\ref{fig:intensity_19Feb2007}.

The Ca $396.8$~nm H-line, shown in the upper right panel of
Figure~\ref{fig:intensity_19Feb2007}, is formed partly in the
photosphere, partly in the chromosphere \citep{Carlsson+etal2007}.
Clearly visible throughout the image is the reverse granulation formed
just above the visible photosphere. Movies made of the Ca H-line show 
the expected evolution of the reverse granulation as well as the
fairly large scale intensity variations due to photospheric p-modes and
chromospheric 3-minute oscillations. Also spread throughout the image are
bright points, corresponding to G-band bright points where one can see
deeper into the solar atmosphere. The bright points associated with
the small magnetic field elements and the plage show motions well
correlated with the motions of the underlying field. 
In addition, especially surrounding
the plage,we find `hazy' regions that most likely image chromospheric
emission at heights far above the photosphere: movies constructed using 
a high-pass frequency filter (retaining only frequencies $>25$~mHz) show 
that this haze corresponds to the `straws' seen with $0.1$~nm wide Ca H-line 
filtergram movies at the Dutch Open Telescope \citep{Rutten2007} and on the solar 
limb with Hinode \citep{DePontieu+etal2007b}. (Straws appear hazy since they are 
nearly straight, short-lived features that are formed at middle to upper
chromospheric heights, and that are weak because of the large
photospheric contribution to the wide Ca H filter used on SOT.) 

\begin{figure}
  \begin{center}
    \FigureFile(80mm,80mm){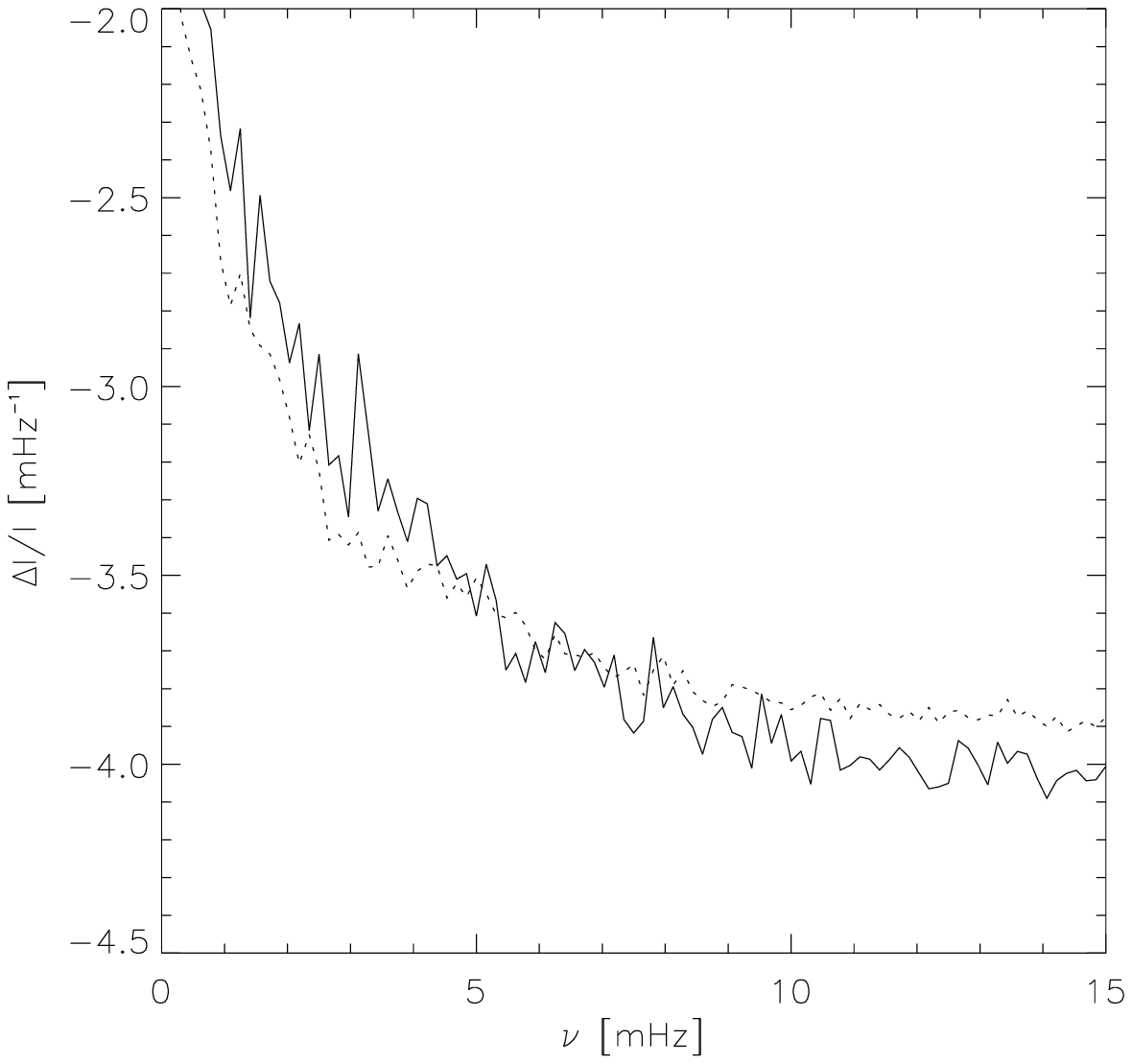}
    \FigureFile(80mm,80mm){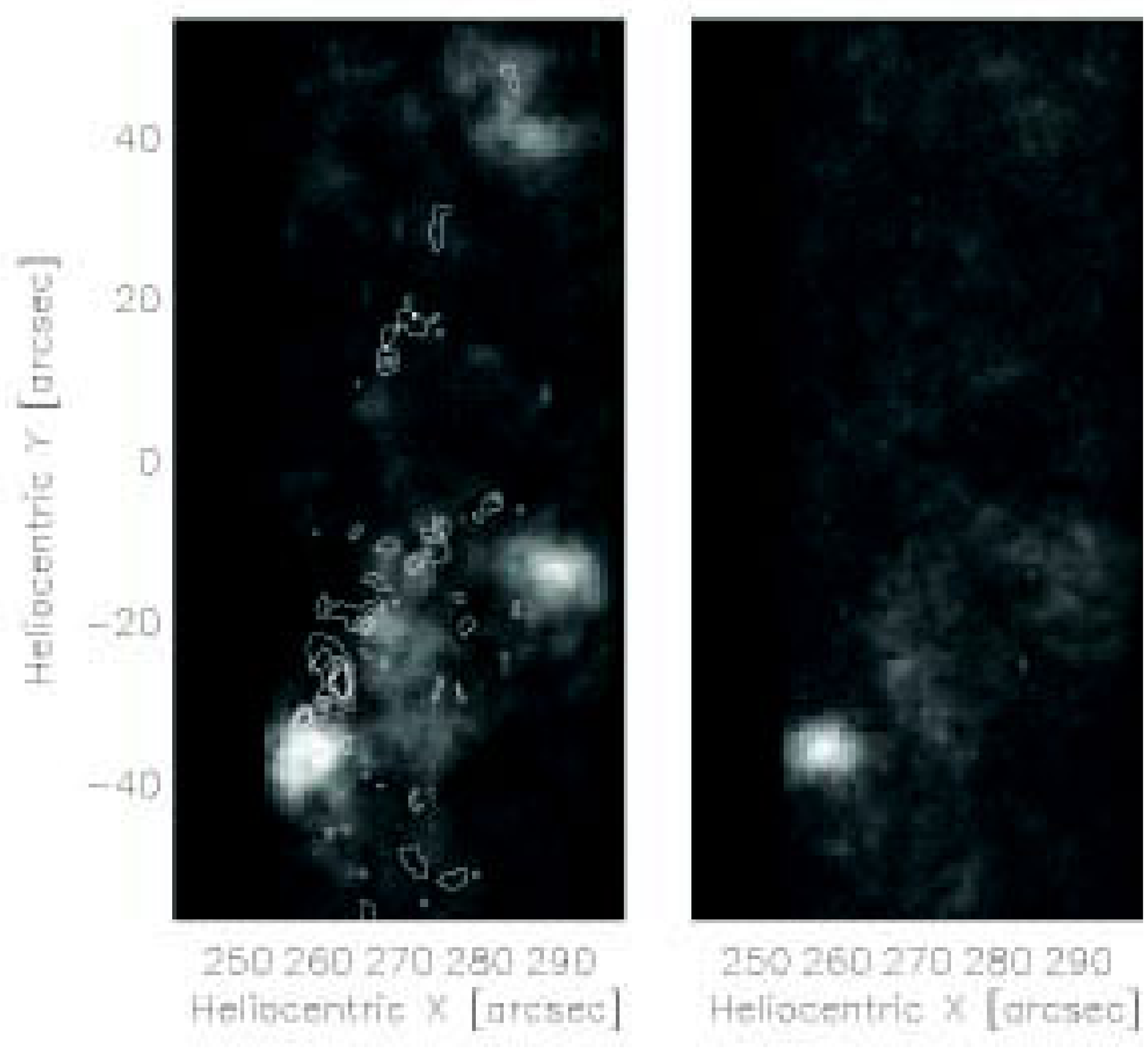}
  \end{center}
  \caption{The upper panel shows the power frequency spectrum for the variation of the 
           logarithm of the intensity $\Delta I/I$ for the He~{\sc ii} $25.6$~nm line
           in the February 19, 2007 quiet Sun data-set 
           for a subset of the brighter pixels found in the vicinity of the weak plage region (solid line) 
           and for a subset of pixels imaging typical quiet Sun regions (dotted line). 
           The lower panels show maps of the location of power in this line in 1~mHz wide 
           bands centered
           on 3~mHz (left panel) and 5~mHz (right panel). Overplotted on the 3~mHz map are
           the contours of the SOT magnetogram shown in Figure~\ref{fig:intensity_19Feb2007}.
           Axes are heliocentric co-ordinates measured in arcseconds.
           }
           \label{fig:power_1920Feb2007}
\end{figure}

As seen in 
Figure~\ref{fig:emergence_19Feb2007} the region
co-spatial with and/or just above the region of flux emergence seen in
the magnetogram movies initially shows a dimming in the Ca H-line
filtergrams. Flux is first seen to emerge in the magnetograms 
at 14:44~UT. This is followed by 
what appear to be two to three large dark granules in the Ca-H-line some 10~minutes later
at 14:55~UT. These dark granules expand
and move with the emergence of the field. The border of this dark region
becomes filled with bright points within 20~minutes of the fields
emergence. As time passes the dark granules brighten, seemingly as a
result of `haze' filling in the dark region, which upon inspection of 
difference and high-pass frequency filtered movies are revealed to consist of loop-like 
structures in the Ca-H-line that join the newly formed bright points.

In the lower left panel of Figure~\ref{fig:intensity_19Feb2007} we show a raster image
of the He {\sc ii} 25.6 nm line which is  formed in the 80,000 K transition region, in the lower
right panel we show the Fe {\sc xii} 19.5 nm coronal line which is formed at some 1~MK. 
In both lines we see the weak plage region, stretching from (255,-35) to (280,-5), delineated 
quite clearly. This is also true for some of the smaller and weaker plage regions to the 
north. On the other hand, the correspondence between bright
emission as seen in the transition region and coronal lines and magnetic flux measured in the
photosphere does not extend all the way down to magnetic
structures of the smallest spatial scales. This lack of correspondence has been observed 
previously in moss regions on spatial scales of 1~arcsec \citep{DePontieu+etal1999}.

EIS 40~arcsec slot images of a number of strong lines, including the 
He~{\sc ii} and Fe~{\sc xii} lines were observed co-temporally with the 
SOT movies described above. These slot movies show large temporal
variations at cadences down to those measurable by the 30~s exposure time;
the general topology outlined by the plage region is maintained, also in the upper
solar atmosphere, but the emission is strongly variable at any given location both within 
the plage and outside in the weaker field quiet Sun. 
These variations 
do not show an obvious one-to-one correspondence with events that
are transpiring below: Some of the magnetic flux cancellations as two
opposite polarity elements coalesce are accompanied by brightening in
the He and Fe lines and some are not. The 3/5 minute wave pattern that
is clearly discernible in the Ca H-line emission does not leave an obvious
footprint in the lines formed above. An exception to this rule of non-correspondence 
is the large flux emergence event that happens in the plage. As is clear by inspection
of Figure~\ref{fig:emergence_19Feb2007} both the He~{\sc ii} and the Fe~{\sc xii} lines 
show an dimming dark ``bubble'' that appears at 15:14~UT and expands thereafter, 
almost 30~minutes after the flux emergence is obvious in the magnetograms. Unfortunately, 
we did not observe the end of this event with EIS as another observing program was 
started at 15:30~UT.

\begin{figure}
  \begin{center}
    \FigureFile(80mm,80mm){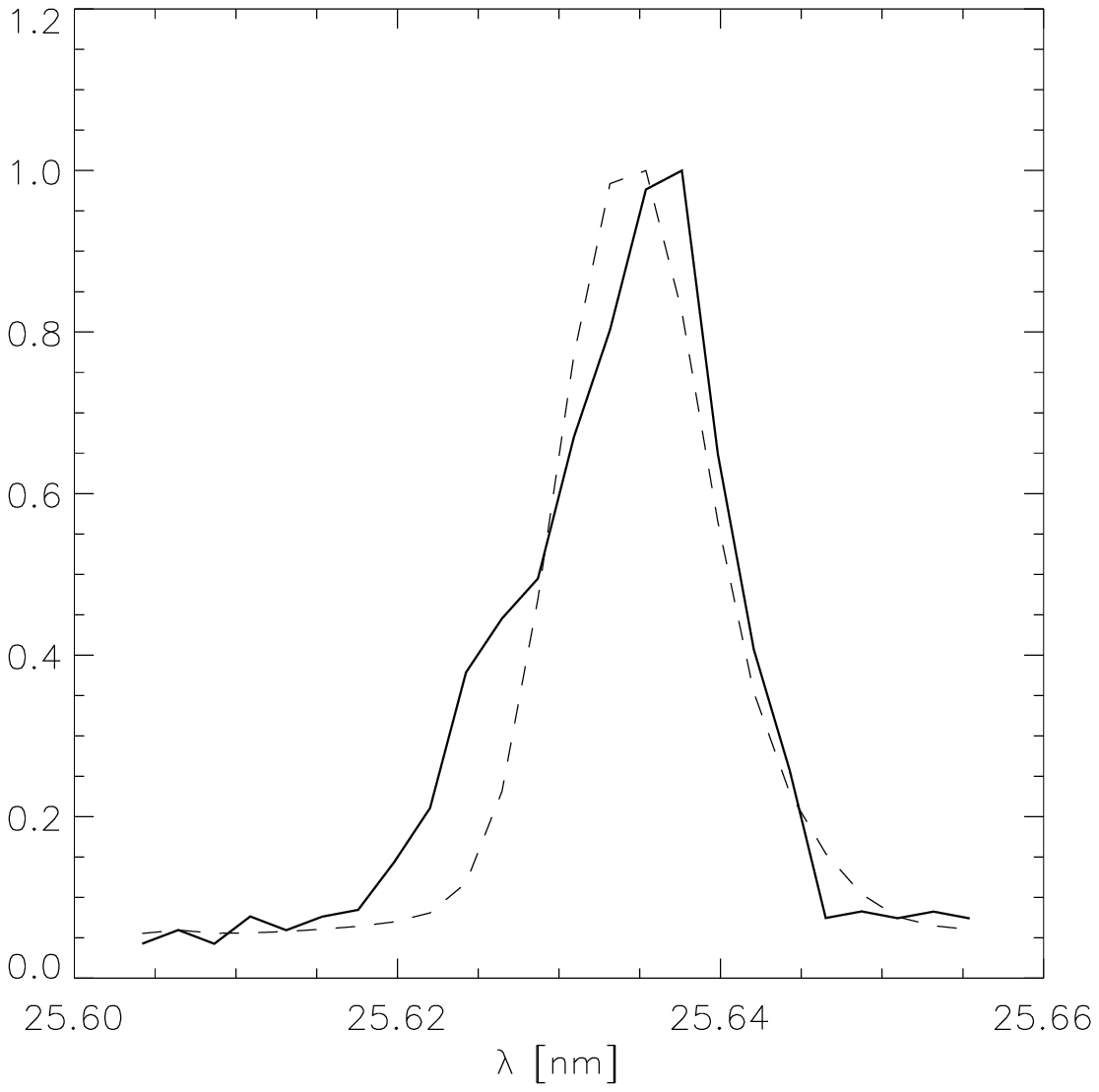}
    \FigureFile(80mm,80mm){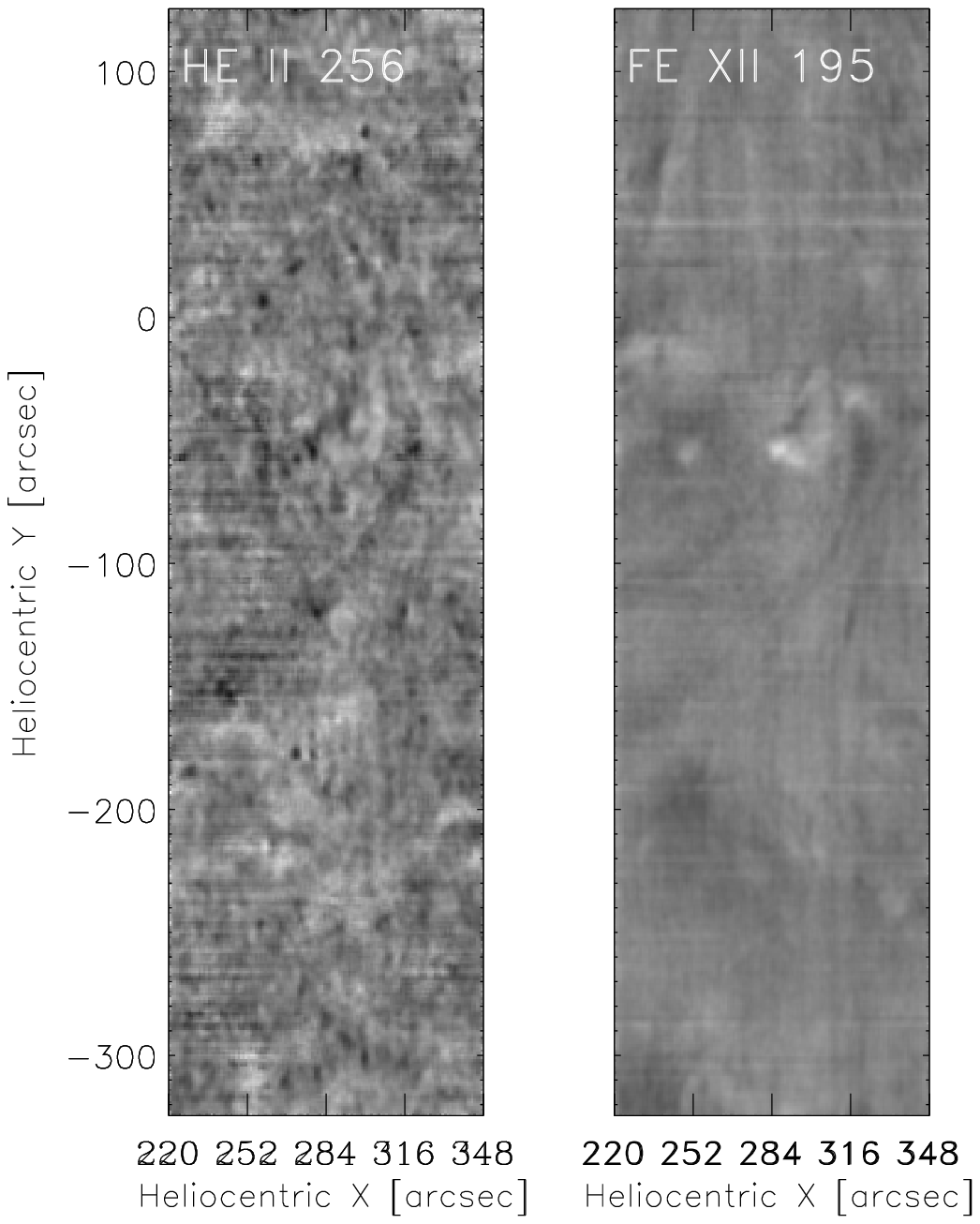}
  \end{center}
  \caption{The upper panel shows line profiles of the He~{\sc ii} 25.6~nm line: 
  Average line profile (dashed line) and a profile showing high velocity event 
  (solid line), as described in the text, and shown in the velocity maps in the
  lower panels, where the high velocity events are evident as small dark dots in the
  image. The right panel shows
  a velocity map of the Fe~{\sc xii} 19.5~nm line. The velocity maps were made from 
  rasters obtained on February 19, 2007 and have a range
  of $\pm 25$~km/s. Axes are heliocentric co-ordinates measured in arcseconds.}
  \label{fig:lineshift_velocity_20Feb2007}
\end{figure}

\subsection{Wave Propagation}

\begin{figure*}
  \begin{center}
    \FigureFile(150mm,70mm){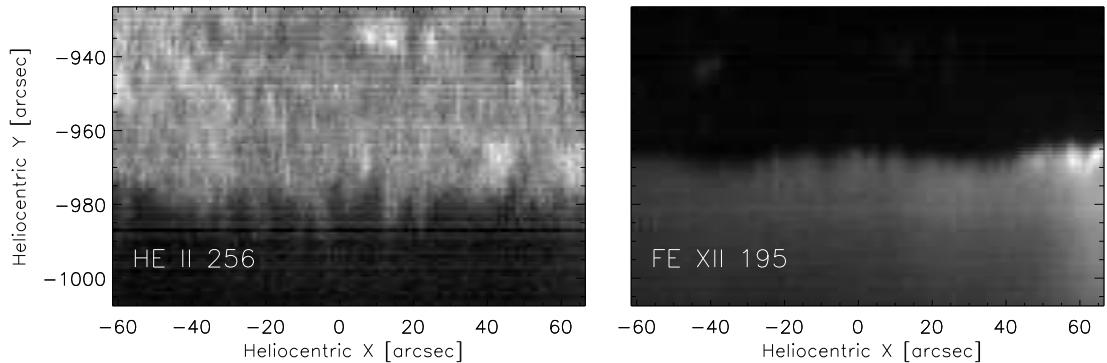}
  \end{center}
  \caption{South limb images made from rasters in the He~{\sc ii} 25.6~nm and Fe~{\sc xii} 19.5~nm lines
  obtained February 16, 2007.
  Note the (macro)-spicules in the He~{\sc ii} line and that similar features appear in absorption
  in the Fe~{\sc xii} line. Axes are heliocentric co-ordinates measured in arcseconds.}
  \label{fig:spicules_16Feb2007}
\end{figure*}

While there were few obvious correlations between the movies made with
SOT filtergrams and the EIS slot, an initial insight into the role of
waves can be found by considering the power contained in the 3- and
5-minute frequency bands of this region in EIS He~{\sc ii} slot movies.
In the upper panel of figure~\ref{fig:power_1920Feb2007} we show the
computed power spectra of the variation of the intensity $\Delta I/I$
for a subset of the brighter pixels found in the vicinity of the plage
region (solid line) and for a subset of pixels imaging typical quiet Sun
regions (dotted line). The spectra show some excess power for the
brighter region, in both 3 and 5 minute bands, while the power found in
the quiet Sun region is of questionable significance. Note however, that
the wave power found in the transition region is known from SUMER
observations ({\it e.g.} \cite{McIntosh+etal2001}) to be quite dependent on the
topology of the magnetic field in the chromosphere below. Significant 5
minute power in coronal lines has been observed before with TRACE, e.g., in
the TRACE 17.1~nm and 19.5~nm bands in loops \citep{DeMoortel+etal2002a, DeMoortel+etal2002b} 
as well as in the moss \citep{DePontieu+etal2003, DePontieu+etal2005}.
Likewise, 3~minute power in coronal emission above active regions has also been 
seen before \citet{2000SoPh..191..129B} in the TRACE 17.1~nm band.

Maps of the location of the Fourier power of the intensity variation
$\Delta I/I$ of the He~{\sc ii} 25.6~nm line were constructed by taking the sum
over time of the absolute value of the Fourier filtered intensity
signal. In the lower panels of Figure~\ref{fig:power_1920Feb2007} 
we show the power maps derived for 1~mHz
wide bands centered on 3~mHz (left panel) and 5~mHz (right panel). We
find the power to be concentrated in the vicinity of the plage as
indicated by the contours of the absolute value of the magnetic field
strength drawn into the 3~mHz map. This could be an example of the
leakage of p-modes into the upper atmosphere along inclined magnetic
field lines ({\it e.g.} \cite{DePontieu+etal2004}). However, identification
of these intensity variations as wave-like as opposed to being a result of 
evolution (of {\it e.g.} the magnetic field) awaits the analysis of EIS velocities 
in sit-and-stare studies with sufficiently high cadence. Future work involving field
extrapolations based on the SOT magnetograms and travel time analysis in
the lower atmosphere will also be necessary to confirm this and/or to pin down the 
exact leakage mechanism.

\subsection{Transition region velocity events}

While our slot movies and intensity raster maps do not show any clear evidence of
small scale features that can be linked with the lower atmosphere,
velocity maps made in the transition region He~{\sc ii} $25.6$~nm line do show small scale structure that
should have some lower lying counterpart.  In the lower panels of 
Figure~\ref{fig:lineshift_velocity_20Feb2007} we present velocity maps of 
the rasters obtained in the vicinity of the small active region NOAA 10942. 

\begin{figure}
  \begin{center}
    \FigureFile(80mm,140mm){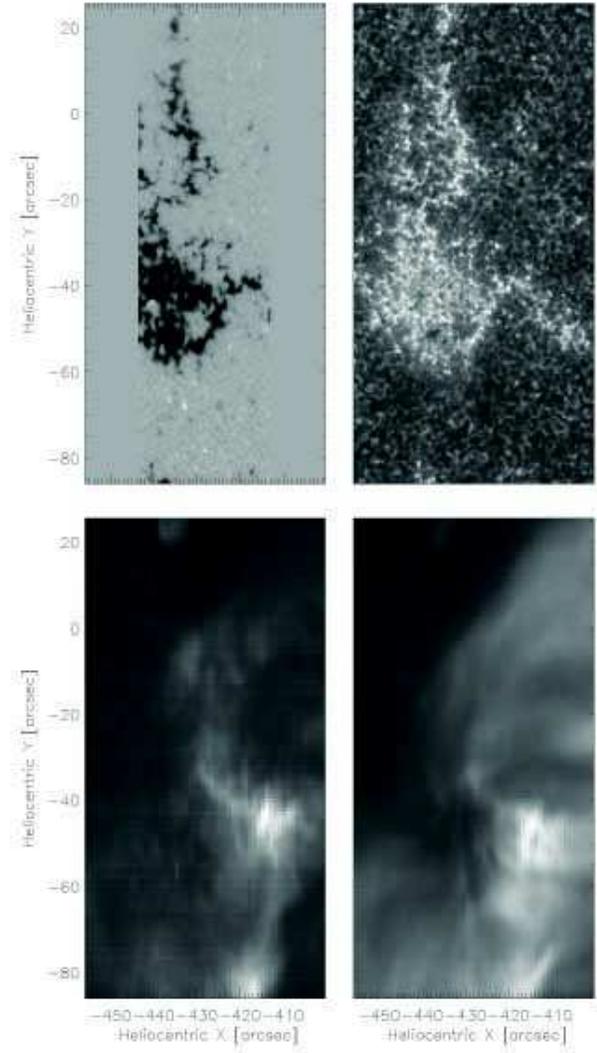}
  \end{center}
  \caption{Co-pointed SOT and EIS images of the small active region NOAA~10942 centered on solar co-ordinates
  (-430,-30) obtained February 20, 2007. In the upper left panel we show a Fe~{\sc i} 630.2~nm 
  magnetogram and in the upper right panel the Ca~{\sc ii} 396.8~nm H-line. In the lower panels
  the corresponding EIS raster images for the transition region He~{\sc ii} 25.6~nm (left) and the
  coronal Fe~{\sc xii} 19.5~nm (right) lines are shown. Axes are heliocentric co-ordinates measured in arcseconds.}
  \label{fig:intensity_20Feb2007}
\end{figure}

The velocity maps are both normalized so as to show line shifts of $\pm 25$~km/s 
relative to the average line shift, 
with black color indicating upward blue-shifts and white indicating down-flowing 
red-shifts. The velocities show a fairly complex structure, and quite striking are
the numerous intense ($>25$~km/s) blue-shifts found throughout the quiet Sun
region of the images. (This type of event is found in every raster made in the 
He~{\sc ii} line that includes quiet Sun emission during the two-week period covered by 
this study). Note that
the blue-shifts are not evident in the 1~MK plasma imaged by the Fe~{\sc xii} line
shown in the lower right panel. The linear extent of these phenomena is less than
1~arcsec$^2$.  Note that that the S~{\sc x} blend known to be located near the He~{\sc ii} 
line is on the {\it red} side of the He~{\sc ii} line (and is visible to the red of
the He~{\sc ii} line in spectra taken above the limb). 

The line profile of a ÔtypicalÕ up-flow event (solid line) along with the
average line profile in the He~{\sc ii} 25.6 nm line is shown in the upper 
panel of Figure~\ref{fig:lineshift_velocity_20Feb2007}. The actual velocity in the
up-flow event is larger than the average velocity derived by taking the first moment
of the line: the upper panel line profile reveals that the line is split into
two components, the component found in the blue wing has a relative shift of some
$100$~km/s. This phenomena may be similar to the events found by 
\citet{Dere+Bartoe+Brueckner1984} with the HRTS instrument.

We are currently examining the SOT magnetograms to see whether we can find any 
correlations between the merging or appearance of magnetic elements and 
these up-flow events seen in He~{\sc ii}, but our first impression is that there is no one-to-one 
correspondence between these events.

\subsubsection{Spicules and macro-spicules}

As an aside, we mention the possibility that there is a connection between 
the up-flow events described here and the He~{\sc ii} (macro)-spicules observed 
above the limb. An example is the EIS raster made on the southern pole limb on February 16 2007, 
shown in Figure~\ref{fig:spicules_16Feb2007}. Similar spicules have been observed previously in 
the EUV passband by SOHO/SUMER \citep{Wilhelm2000}. The observed ``spicules'' appear as
largely radial features extending some 10~arcsec above the limb with widths of order
$1-2$~arcsec ({\it i.e.} roughly the resolution of EIS).  The spicules show up 
in {\it absorption} in the Fe~{\sc xii} line: The bulk of the plasma in this phenomena 
seems not to be heated to coronal temperatures. Note that the spicules are 
quite numerous, they are also continually present in slot movies made on the same date, but 
the lower resolution of the slot mode hinders accurate identification and the measurement
of lifetimes, birthrates or other properties, nor indeed correlation with the Ca~H-line 
spicules found above the limb.

\begin{figure*}
  \begin{center}
    \FigureFile(140mm,80mm){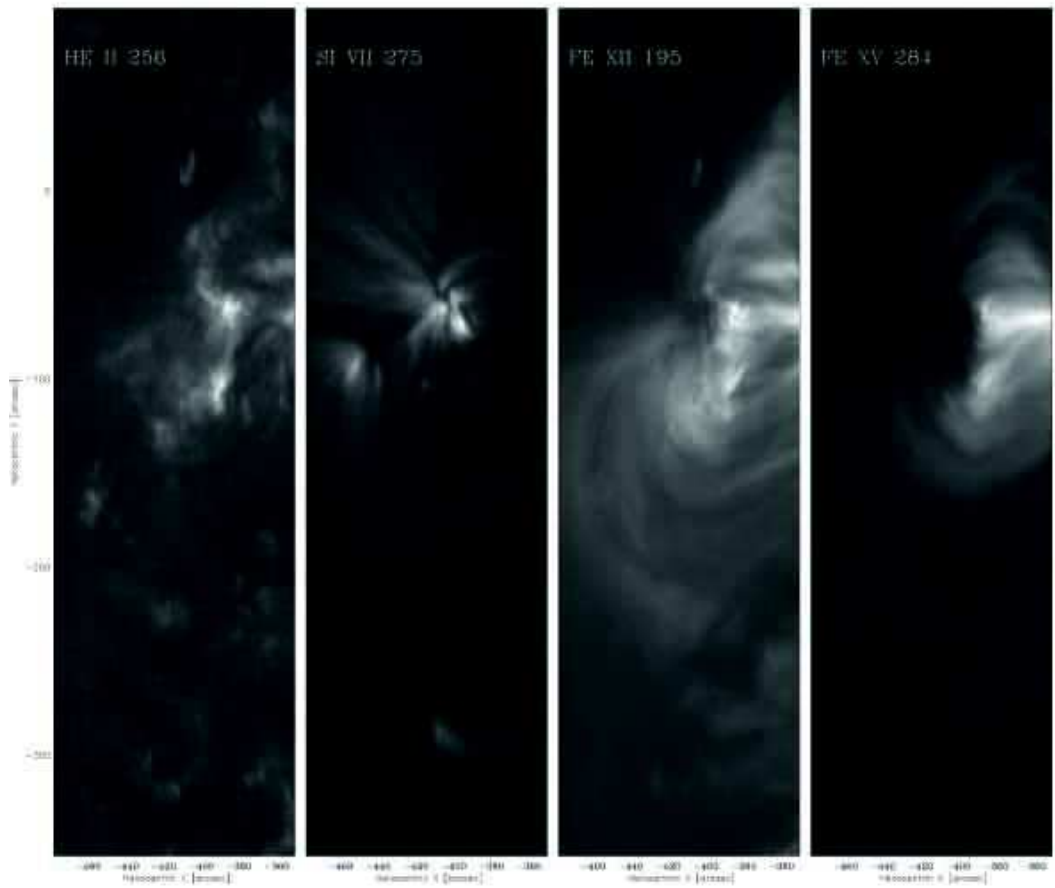}
  \end{center}
  \caption{Intensity maps of NOAA 10942 made from EIS rasters on February 20, 2007. From left to
  right the lines shown are He~{\sc ii} 25.6~nm, Si~{\sc vii} 27.5~nm, Fe~{\sc xii} 19.5~nm and Fe~{\sc xv} 28.4~nm.
  Axes are heliocentric co-ordinates measured in arcseconds.}
  \label{fig:intensity_all_20Feb2007}
\end{figure*}

\subsection{NOAA 10942}

The difficulties in connecting phenomena in the lower solar atmosphere
with the transition region and corona is driven home by the extended set
observations made of NOAA 10942 made on February 20 (and every day
thereafter until February 27). In the photosphere this region displays a
large aggregate of (largely) unipolar field as shown in the upper left 
panel of Figure~\ref{fig:intensity_20Feb2007}. The Ca H-line images, upper right
panel, show enhanced emission in the plage region, presumably both from the 
photosphere and from the chromosphere above. A small pore is apparent 
towards the lower part of the plage. 

It is very difficult to pick out the structure
of the plage region in images taken of the hotter plasma above, as should be clear from 
the He~{\sc ii} 25.6~nm and Fe~{\sc xii}~19.5~nm images shown in the lower panels of 
Figure~\ref{fig:intensity_20Feb2007} as well as from the larger field of view images
shown in Figure~\ref{fig:intensity_all_20Feb2007}. The brighter regions in the transition
region and coronal lines lie somewhat to the east of the plage region and most of the
hotter loops seen in Fe~{\sc xii} extend eastward, apparently rooted in the western
edge of the plage region. Note that also all the rasters made with the EIS instrument, 
including those shown in Figure~\ref{fig:intensity_all_20Feb2007} differ
markedly from each other. The He~{\sc ii} emission seems to mainly consist of shorter loops
and perhaps hints of chromospheric network (though the correlation with the Ca~H-line emission
is poor in this example). The Si~{\sc vii} 27.5~nm line, formed in the upper transition region 
at 630~kK, shows emission in the plage region offset slightly to the west compared to He~{\sc ii} and 
Fe~{\sc xii}. The loops in this line mainly extend north-westward across the plage, the 
eastward directed loops are partial, presumably footpoints of the hotter, eastward oriented, 
loops we find in Fe~{\sc xii} and in the Fe~{\sc xv} 28.4~nm line formed at some $2.1$~MK.

\section{Conclusions}

Both differences in instrument temporal and spatial resolution as well as fundamental 
differences between the photosphere/chromosphere and the outer solar layers comprising
the transition region and corona are significant stumbling blocks in understanding the 
coupling between these regions. Even though the Hinode spacecraft presents us with tools 
of unprecedented sophistication for unravelling the complex of physical processes that 
control the outer layers of the Sun, a successful result will require using advanced analysis
techniques. It is fortunate that these techniques are entering the scene concurrently with the 
observations so as to make comparison between theory and observation meaningful.

Hinode is a Japanese mission developed and launched 
by ISAS/JAXA, collaborating with NAOJ as domestic 
partner, NASA and STFC (UK) as international partners. 
Scientific operation of the Hinode mission is conducted 
by the Hinode science team organized at ISAS/JAXA. 
This team mainly consists of scientists from institutes in 
the partner countries. Support for the post-launch operation is provided by JAXA and NAOJ (Japan), STFC 
(U.K.), NASA (U.S.A.), ESA, and NSC (Norway). This work was supported 
by the Norwegian Research Council grant 170926.


\end{document}